\documentclass[twocolumn,preprintnumbers,amsmath,amssymbm,prl]{revtex4}
\usepackage{epsfig}
\usepackage{graphicx}

\begin{document}
\title{Analytic treatment of the network synchronization problem with time delays}
\author{Shahar Hod}
\affiliation{The Ruppin Academic Center, Emeq Hefer 40250, Israel}
\affiliation{ } \affiliation{The Hadassah Institute, Jerusalem
91010, Israel}
\date{\today}

\begin{abstract}
\ \ \ Motivated by novel results in the theory of network
synchronization, we analyze the effects of nonzero time delays in
stochastic synchronization problems with linear couplings in an
arbitrary network. We determine {\it analytically} the fundamental
limit of synchronization efficiency in a noisy environment with
uniform time delays. We show that the optimal efficiency of the
network is achieved for
$\lambda\tau={{\pi^{3/2}}\over{2\sqrt{\pi}+4}}\approx0.738$, where
$\lambda$ is the coupling strength (relaxation coefficient) and
$\tau$ is the characteristic time delay in the communication between
pairs of nodes. Our analysis reveals the underlying mechanism
responsible for the trade-off phenomena observed in recent numerical
simulations of network synchronization problems.
\end{abstract}
\bigskip
\maketitle


Synchronization processes in populations of locally interacting
elements are in the focus of intense research in physical,
biological, chemical, technological and social systems
\cite{Arenas}. Of particular interest are situations in which
members (usually referred to as `agents' or `nodes' in a network)
try to coordinate their state in a {\it decentralized} manner
\cite{Arenas,HKS}. In many real-life situations the motivation for
such coordination is to improve the global performance of the
network \cite{HKS}. There has been a flurry of research focusing on
the efficiency and optimization of synchronization problems in
various complex network topologies (see
\cite{Arenas,HKS,Kor1,Kor2,Bara,Nish1,Atay,La,Zhou,Nish2,Olf,Hod1,Hod2}
and references therein).

Stochastic synchronization problems in real biological, social, and
computing networks are usually characterized by finite {\it time
delays} in the communication between pairs of nodes. Recently, Hunt
{\it et al.} \cite{HKS} have studied the impact of such time delays
on synchronizability and on the breakdown of synchronization in
dynamical network-connected systems. They considered a stochastic
model in which each node in a network adjusts its state to match
that of its neighbors, but with a uniform time lag in reacting to
the neighborly feedback. Hunt {\it et al.} have revealed that there
are trade-offs in the synchronization problem: when there are large
lag times in communication between nodes, reduced local coordination
effort may actually improve the global coordination of the network
\cite{HKS}.

It is worth emphasizing that the remarkable finding of \cite{HKS},
that there are possible scenarios for trade-offs between large time
lags and the coupling strength, is based on {\it numerical}
simulations of the stochastic evolution equations which govern the
dynamics of the network [see Eq. (\ref{Eq6}) below]. The main goal
of the present Letter is to provide an {\it analytical} treatment
for the network synchronization problem. In particular, we shall
determine analytically the fundamental limit of synchronization
efficiency in a noisy environment with uniform time delays.

We shall first describe the synchronization model studied in Ref.
\cite{HKS}. Consider a stochastic model where $N$ agents in a
network locally adjust their state in an attempt to match that of
their neighbors. Such coordination may improve the global
performance of the network \cite{Arenas,HKS,Bara,Nish1,La}. As in
many real-life situations, the communication between pairs of nodes
is not instantaneous. Rather, it is characterized by some finite
time lag \cite{HKS}. The dynamics of the system is governed by the
coupled stochastic equations of motion with linear local relaxation
and a uniform time delay,
\begin{equation}\label{Eq1}
{{\partial h_i(t)}\over{\partial t}}=-\sum_{j=1}^N
C_{ij}[h_i(t-\tau)-h_j(t-\tau)]+\eta_i(t)\  ,
\end{equation}
where $h_i(t)$ is the generalized local state variable on node $i$,
$C_{ij}=C_{ji}\geq0$ is the symmetric coupling strength between two
connected nodes $i$ and $j$, and $\tau$ is the characteristic time
delay between two connected nodes. Here $\eta_i(t)$ is a
delta-correlated noise with zero mean and variance
$\langle\eta_i(t)\eta_j(t')\rangle=2D\delta_{ij}\delta(t-t')$, where
$D$ is the noise intensity \cite{HKS}.

Stochastic synchronization problems are characterized by competition
between a relaxation mechanism and a random noise. The physically
interesting observable in such systems is the width of the
synchronization landscape. This is given by \cite{HKS,Kor1,Kor2,La}
\begin{equation}\label{Eq2}
\langle{w^2(t)}\rangle\equiv\Big\langle{1\over
N}\sum_{i=1}^N[h_i(t)-\bar h(t)]^2\Big\rangle\ ,
\end{equation}
where $\bar h(t)={1/N}\sum_{i=1}^N h_i(t)$ is the global average of
the local state variables and $\langle\cdot\cdot\cdot\rangle$
denotes an ensemble average over the noise. A network is considered
synchronizable if its late-time asymptotic behavior is characterized
by a {\it finite} width [that is, if $\langle
w^2(\infty)\rangle<\infty$]. The smaller the width, the better the
synchronization \cite{HKS}.

The coupled equations of motion (\ref{Eq1}) can be rewritten as
\cite{HKS}
\begin{equation}\label{Eq3}
{{\partial h_i(t)}\over{\partial t}}=-\sum_{j=1}^N
\Gamma_{ij}h_j(t-\tau)+\eta_i(t)\  ,
\end{equation}
where $\Gamma_{ij}=\delta_{ij}\sum_lC_{il}-C_{ij}$ is the symmetric
network Laplacian. Further, by diagonalizing the network Laplacian,
one can decompose the problem into $N$ {\it independent} modes
\begin{equation}\label{Eq4}
{{\partial \tilde h_k(t)}\over{\partial t}}=-\lambda_k \tilde
h_k(t-\tau)+\tilde\eta_k(t)\  ,
\end{equation}
where $\{\lambda_k\}$ ($k=0,1,2,...,N-1$) are the eigenvalues of the
network Laplacian and
$\langle\tilde\eta_k(t)\tilde\eta_l(t')\rangle=2D\delta_{kl}\delta(t-t')$.
For a connected (single-component) network, the Laplacian has a
single zero mode (indexed by $k=0$) with $\lambda_0=0$, while
$\lambda_k>0$ for $k\geq1$ \cite{HKS}. Using the above eigenmode
decomposition, the width of the synchronization landscape can be
expressed as \cite{HKS}
\begin{equation}\label{Eq5}
\langle w^2(t)\rangle={1\over N}\sum_{k=1}^{N-1}\langle\tilde
h^2_k(t)\rangle\ .
\end{equation}

Note that the eigenmodes of the system are governed by a stochastic
equation of motion [Eq. (\ref{Eq4})] of {\it identical} form for all
$k\geq1$. We shall therefore omit the index $k$ for brevity, and
study the stochastic differential equation
\begin{equation}\label{Eq6}
{{\partial \tilde h(t)}\over{\partial t}}=-\lambda \tilde
h(t-\tau)+\tilde\eta(t)\
\end{equation}
with $\langle\eta(t)\eta(t')\rangle=2D\delta(t-t')$.

Using a Laplace transformation with initial conditions $\tilde
h(t\leq0)=0$, one finds \cite{HKS}
\begin{equation}\label{Eq7}
\tilde h(t)=\int_0^t dt'\tilde\eta(t')
\sum_{\alpha}{{e^{s_{\alpha}(t-t')}}\over{1+\tau s_{\alpha}}}\  ,
\end{equation}
where $\{s_{\alpha}\}$ ($\alpha=1,2,...$) are the solutions of the
characteristic equation
\begin{equation}\label{Eq8}
s+\lambda e^{-\tau s}=0\
\end{equation}
in the complex plane. The characteristic equation (\ref{Eq8}) has an
infinite number of complex solutions for $\tau>0$
\cite{Olf,Fris,Hay}. In particular, it is well known that
$\Re(s_{\alpha})<0$ for all $\alpha$ provided $\lambda\tau<\pi/2$
\cite{Olf,Fris,Hay}.

Using Eq. (\ref{Eq7}), one finds \cite{HKS}
\begin{equation}\label{Eq9}
\langle\tilde
h^2(t)\rangle=\sum_{\alpha}\sum_{\beta}{{-2D\tau[1-e^{(z_{\alpha}+z_{\beta})t/\tau}]}\over{(1+z_{\alpha})(1+z_{\beta})(z_{\alpha}+z_{\beta})}}\
\end{equation}
for the noise-averaged fluctuations, where $z\equiv\tau s$.
Inspection of Eq. (\ref{Eq9}) reveals that the condition for
$\langle\tilde h^2(\infty)\rangle$ to remain finite is
$\Re(z_{\alpha})<0$ for all $\alpha$. As discussed above, this
requires $\lambda\tau<\pi/2$ \cite{Note1}. In the synchronizable
regime [$\Re(z_{\alpha})<0$ for all $\alpha$] one finds
\begin{equation}\label{Eq10}
\langle\tilde
h^2(\infty)\rangle=\sum_{\alpha}\sum_{\beta}{{-2D\tau}\over{(1+z_{\alpha})(1+z_{\beta})(z_{\alpha}+z_{\beta})}}\
\end{equation}
for the steady-state ($t\to\infty$) behavior.

Writing the characteristic equation (\ref{Eq8}) in the form
\begin{equation}\label{Eq11}
z+\lambda\tau e^{-z}=0\  ,
\end{equation}
one realizes that $z_{\alpha}=z_{\alpha}(\lambda\tau)$ \cite{HKS}.
Thus, one immediately deduces from Eq. (\ref{Eq10}) the scaling form
\begin{equation}\label{Eq12}
\langle\tilde h^2(\infty)\rangle=D\tau\times f(\Lambda)\  ,
\end{equation}
where $\Lambda\equiv\lambda\tau$. The scaling function $f(\Lambda)$
was constructed {\it numerically} in \cite{HKS}. In particular, the
numerical study of $f(\Lambda)$ in \cite{HKS} yielded the remarkable
finding that $f(\Lambda)$ is a {\it non}-monotonic function; it
exhibits a single minimum, at approximately $\Lambda^*\approx0.73$
with $f(\Lambda^*)\approx3.1$ (see Fig. 2 of \cite{HKS}).

Our main goal here is to provide an {\it analytical} treatment for
the problem of network synchronization in a noisy environment with
time delays. To that end, we shall first analyze the asymptotic
behavior of $\langle\tilde h^2(\infty)\rangle$ near the two
boundaries of the synchronizable regime: $\Lambda\to0$ and
$\Lambda\to\pi/2$. As we shall show below, in these limits the sum
in (\ref{Eq10}) is dominated by solutions of the characteristic
equation (\ref{Eq11}) with $\Re(z)\to0$.

In the $\Lambda\to0$ limit, the function $f(\Lambda)$ has to scale
as
\begin{equation}\label{Eq13}
f(\Lambda\to0)\simeq {{1}\over{\Lambda}}+O(1)\
\end{equation}
in order to reproduce the exact limiting case of zero delay,
$\langle\tilde h^2(\infty)\rangle\simeq D/\lambda$ \cite{HKS}.

In the $\Lambda\to\pi/2$ limit we find the pair of solutions
\begin{equation}\label{Eq14}
z_{\pm}=\pm i{{\pi}\over{2}}-{{\pm
i+{{\pi}\over{2}}}\over{1+{({\pi\over2})^2}}}\Delta+O(\Delta^2)\
\end{equation}
to the characteristic equation (\ref{Eq11}), where
$\Delta\equiv\pi/2-\Lambda\ll1$. Note that
\begin{equation}\label{Eq15}
z_++z_-=-{{\pi}\over{1+{({\pi\over2})^2}}}\Delta\to0
\end{equation}
in the $\Delta\to0$ limit. Inspection of the denominator of Eq.
(\ref{Eq10}) reveals that the small value of the sum $z_++z_-$ is
responsible for the divergent behavior of $f(\Lambda\to\pi/2)$.
Substituting (\ref{Eq15}) into Eq. (\ref{Eq10}), one obtains the
leading divergent behavior of $f(\Lambda)$ in the
$\Lambda\to{{\pi}\over{2}}$ ($\Delta\to0$) limit:
\begin{equation}\label{Eq16}
f(\Lambda\to{{\pi}\over{2}})\simeq {{4}\over{\pi\Delta}}+O(1)\  .
\end{equation}

The {\it simplest} analytic function which satisfies both asymptotic
behaviors (\ref{Eq13}) and (\ref{Eq16}) is
\begin{equation}\label{Eq17}
f(\Lambda)={{1}\over{\Lambda}}+{{4}\over{\pi({{\pi}\over{2}}-\Lambda)}}+c\
,
\end{equation}
where $c$ is a constant. Note that this function has a single
minimum at
\begin{equation}\label{Eq18}
\Lambda^*={{\pi^{3/2}}\over{2(\sqrt{\pi}+2)}}\  .
\end{equation}
We note that the {\it numerically} computed value
$\Lambda^*\approx0.73$ \cite{HKS} is astonishingly close ($\sim1\%$
difference) to the {\it analytical} expression (\ref{Eq18}).

In order to fix the value of the constant $c$ in (\ref{Eq17}), one
may calculate the sub-leading (constant) term in Eq. (\ref{Eq13}).
In the $\Lambda\to0$ limit we find the solution
\begin{equation}\label{Eq19}
z_0=-\Lambda-\Lambda^2+O(\Lambda^3)\
\end{equation}
to the characteristic equation (\ref{Eq11}). Inspection of the
denominator of Eq. (\ref{Eq10}) reveals that the small value of
$z_0$ is responsible for the divergent behavior of $f(\Lambda\to0)$.
Substituting (\ref{Eq19}) into Eq. (\ref{Eq10}), one obtains the
leading divergent behavior of $f(\Lambda)$ in the $\Lambda\to0$
limit:
\begin{equation}\label{Eq20}
f(\Lambda\to0)\simeq {{1}\over{\Lambda}}+1\  .
\end{equation}
Equating Eqs. (\ref{Eq17}) and (\ref{Eq20}) for $\Lambda\to0$, one
finds $c=1-{{8}/{\pi^2}}$, which implies
\begin{equation}\label{Eq21}
f(\Lambda)={{1}\over{\Lambda}}+{{4}\over{\pi({{\pi}\over{2}}-\Lambda)}}+1-{{8}\over{\pi^2}}\
\end{equation}
for the scaling function in (\ref{Eq12}).

Substituting $\Lambda^*$ from (\ref{Eq18}) into (\ref{Eq21}), one
obtains the minimal value
\begin{equation}\label{Eq22}
f_{\text{min}}=f(\Lambda^*)=1+2\pi^{-1}+8\pi^{-3/2}\  .
\end{equation}
Again, we note that the {\it numerically} computed value
$f_{\text{min}}\approx3.1$ \cite{HKS} is remarkably close ($\sim1\%$
difference) to the {\it analytical} expression (\ref{Eq22}).

\begin{figure}[h]
  \begin{center}
    \epsfxsize=8.5cm \epsffile{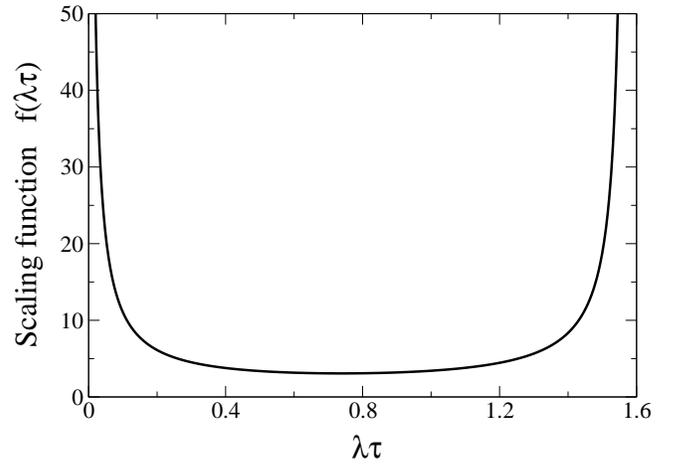}
  \end{center}
  \caption{The scaling function $f(\Lambda)\equiv\langle\tilde
  h^2(\infty)\rangle/D\tau=\Lambda^{-1}+{{4}\over{\pi}}({{\pi}\over{2}}-\Lambda)^{-1}+1-{{8}\over{\pi^2}}$
  in the synchronizable regime $0<\Lambda<\pi/2$. Compare this figure with the {\it numerically} constructed
  function presented in Fig. 2 of \cite{HKS}.}
  \label{Fig1}
\end{figure}

In figure \ref{Fig1} we depict the scaling function
$f(\Lambda)=\langle\tilde h^2(\infty)\rangle/D\tau$ as given by Eq.
(\ref{Eq21}). This figure should be compared with the numerical
results presented in Fig. 2 of \cite{HKS}. We find an almost perfect
agreement between the analytical function (\ref{Eq21}) and the
numerical results of Ref. \cite{HKS}.

\input{epsf}

From Eqs. (\ref{Eq18}) and (\ref{Eq22}) one learns that for a single
stochastic variable governed by Eq. (\ref{Eq6}) with a nonzero
delay, there is an optimal value of the relaxation coefficient
$\lambda^*={{\pi^{3/2}}/{2(\sqrt{\pi}+2)}}\tau$, at which point the
steady-state fluctuations attain their minimum value $\langle\tilde
h^2(\infty)\rangle_{\text{min}}=D\tau (1+2\pi^{-1}+8\pi^{-3/2})$
\cite{Note2}, see also \cite{HKS}.

Returning to the context of network synchronization, one can
calculate from Eqs. (\ref{Eq5}),\ (\ref{Eq12}) and (\ref{Eq21}) the
steady-state width of the network-coupled system:
\begin{equation}\label{Eq23}
\langle w^2(\infty)\rangle={{D\tau}\over{N}}\sum_{k=1}^{N-1}
\Big[{{1}\over{\lambda_k\tau}}+{{4}\over{\pi({{\pi}\over{2}}-\lambda_k\tau)}}+1-{{8}\over{\pi^2}}\Big]\
.
\end{equation}
Thus, for large $N$ the fundamental limit of synchronization
efficiency is given by [see Eq. (\ref{Eq22})]:
\begin{eqnarray}\label{Eq24}
\langle
w^2(\infty)\rangle_{\text{min}}=D\tau(1+2\pi^{-1}+8\pi^{-3/2})\ .
\end{eqnarray}
This is the {\it minimum} attainable width of the synchronization
landscape in a noisy environment with uniform time delays.

So far we have studied the characteristics of the synchronization
network in the steady state ($t\to\infty$) regime. Another
interesting characteristic of the synchronization problem is the
relaxation time of the network, the time it takes for the system to
relax to its finite steady-state width (in the synchronizable
regime, $0<\lambda\tau<\pi/2$). As we shall now show, this
relaxation time diverges in the $\lambda\tau\to\pi/2$ limit, where
the system undergoes a phase transition from a synchronizable state
to an unsynchronizable state.

Taking cognizance of Eq. (\ref{Eq9}), one realizes that the
relaxation phase of the network (into a steady state behavior) is
governed by the solution of the characteristic equation (\ref{Eq11})
with the largest real part [Since $\Re(\alpha)<0$ in the
synchronizable regime, this amounts to the solution of Eq.
(\ref{Eq11}) with the {\it smallest} absolute value of the real
part.] Inspection of Eq. (\ref{Eq9}) reveals that the characteristic
relaxation time, $T(\lambda,\tau)$ \cite{Note3}, is given by
\begin{equation}\label{Eq25}
T\equiv{{\tau}\over{2\text{min}\{|\Re(z_{\alpha})|\}}}\  .
\end{equation}
Taking cognizance of Eq. (\ref{Eq14}), one finds
\begin{equation}\label{Eq26}
T_{\Delta}=\tau{{1+({{\pi}\over{2}})^2}\over{\pi\Delta}}\  ,
\end{equation}
for the diverging relaxation time of the coupled network in the
vicinity of the phase transition (the $\Delta\to0$ regime).

Further, inserting the pair of solutions $\{z_+,z_-\}$ from
(\ref{Eq14}) into (\ref{Eq9}), one obtains the late-time behavior of
the network near the phase transition:
\begin{eqnarray}\label{Eq27}
\langle\tilde h^2(t)\rangle\simeq \langle\tilde h^2(\infty)\rangle
-{{4D\tau}\over{\pi\Delta}}e^{-t/T_{\Delta}}\Big\{1+{{\Delta}\over{[1+({{\pi}\over{2}})^2]^2}}
\nonumber\\
\times\Big[[({{\pi}\over{2}})^2-1]\sin({{\pi t}/{\tau}})+\pi
\cos({{\pi t}/{\tau}})\Big]\Big\}\  .
\end{eqnarray}
 We thus find that the approach of the network to a steady-state
behavior is characterized by damped temporal oscillations of period
$2\tau$ and a characteristic lifetime $T_{\Delta}$. It is worth
noting that these characteristic oscillations are clearly visible in
the numerical results of Hunt {\it et al.} \cite{HKS} (see Fig. 1 of
\cite{HKS}. Observe, in particular, the temporal oscillations in the
plots for $\lambda\tau=1.5$ and $\lambda\tau=1.6$ which are near the
threshold value $\lambda\tau=\pi/2$ of the phase transition
\cite{Note4}).

In summary, in this Letter we have analyzed the problem of network
synchronization in a noisy environment with uniform time delays
\cite{HKS}. In particular, we have determined {\it analytically} the
fundamental limit of synchronization efficiency (the minimum
attainable value of the width of the synchronization landscape):
$\langle
w^2(\infty)\rangle_{\text{min}}=D\tau(1+2\pi^{-1}+8\pi^{-3/2})$,
where $\tau$ is the characteristic time delay in the communication
between pairs of nodes. We have shown that the optimal efficiency of
the network is achieved for
$\lambda\tau={{\pi^{3/2}}\over{2\sqrt{\pi}+4}}$, where $\lambda$ is
the relaxation coefficient (coupling strength). These analytical
results are in perfect agreement with the recent numerical results
of Ref. \cite{HKS}. Further, we have analyzed the relaxation time of
the network and showed that it diverges in the threshold limit
$\lambda\tau\to\pi/2$.

Our results provide a direct analytical explanation for the
intriguing trade-off phenomena (between the time delay $\tau$ and
the coupling strength $\lambda$) observed in recent numerical
simulations \cite{HKS} of stochastic synchronization problems with
time delays.

\bigskip
\noindent {\bf ACKNOWLEDGMENTS}

This research is supported by the Meltzer Science Foundation. I
thank Oded Hod, Yael Oren and Arbel M. Ongo for helpful discussions.


\end{document}